\documentclass[sigconf]{acmart}

\AtBeginDocument{%
  \providecommand\BibTeX{{%
    \normalfont B\kern-0.5em{\scshape i\kern-0.25em b}\kern-0.8em\TeX}}}

\copyrightyear{2022}
\acmYear{2022}
\setcopyright{acmcopyright}\acmConference[ESEM '22]{ACM / IEEE International Symposium on Empirical Software Engineering and Measurement (ESEM)}{September 19--23, 2022}{Helsinki, Finland}
\acmBooktitle{ACM / IEEE International Symposium on Empirical Software Engineering and Measurement (ESEM) (ESEM '22), September 19--23, 2022, Helsinki, Finland}
\acmPrice{15.00}
\acmDOI{10.1145/3544902.3546235}
\acmISBN{978-1-4503-9427-7/22/09}

\usepackage[shortlabels]{enumitem}
\usepackage{hyperref}

\def\fig {Figure~}

\def\tbl {Table~}

\def\sec {Section~}

\newcommand{\dv}[1]{\href{https://cwe.mitre.org/data/definitions/#1.html}}
\newcommand{\dva}[1]{\href{https://www.cvedetails.com/cwe-details/#1}}
\newcommand{\soa}[1]{\href{https://stackoverflow.com/a/#1}{$A_{#1}$}}

\newlength\WIDTHOFBAR
\usepackage{booktabs}
\usepackage{xcolor}
\usepackage{calc}
\newcommand{\nd}{\vspace{1mm}\noindent}
\definecolor{ao(english)}{rgb}{0.0, 0.5, 0.0}

\newif\ifpienumberinlegend

\begin{document}

\title{Does Collaborative Editing Help Mitigate Security Vulnerabilities in Crowd-Shared IoT Code Examples?}

\author{Madhu Selvaraj}
\affiliation{%
  \institution{University of Calgary}
  \country{Canada}
}
\email{madhumitha.selvaraj@ucalgary.ca}

\author{Gias Uddin}
\affiliation{%
  \institution{University of Calgary}
  \country{Canada}
}
\email{gias.uddin@ucalgary.ca}

\renewcommand{\shortauthors}{Selvaraj and Uddin}

\begin{abstract}
\textbf{\textit{Background}}: With the proliferation of crowd-sourced developer forums,
Software developers are increasingly sharing more coding solutions to programming problems with others in forums.
The decentralized nature of knowledge sharing on sites has raised the concern of sharing security vulnerable code, which then can be reused into mission critical software systems - making those systems vulnerable in the process. Collaborative editing has been introduced in forums like Stack Overflow to improve the quality of the shared contents.
\textbf{\textit{Aim}}: In this paper, we investigate whether code editing can mitigate shared vulnerable code examples by analyzing IoT code snippets and their
revisions in three Stack Exchange sites: Stack Overflow, Arduino, and Raspberry Pi.
\textbf{\textit{Method}}:We analyze the vulnerabilities present in shared IoT C/C++ code snippets, as C/C++ is one of the most widely used languages in mission-critical devices and low-powered IoT devices. We further analyse the revisions made to these code snippets, and their effects.
\textbf{\textit{Results}}: We find several vulnerabilities such as \dv{788}{CWE 788 - Access of Memory Location After End of Buffer}, in 740 code snippets .
However, we find the vast majority of posts are not revised, or revisions
are not made to the code snippets themselves (598 out of 740). We also find
that revisions are most likely to result in no change to the number of vulnerabilities in a code snippet rather than deteriorating or improving the snippet.
\textbf{\textit{Conclusions}}: We conclude that the current collaborative editing system in the forums
may be insufficient to help mitigate vulnerabilities in the shared code.
\end{abstract}



\maketitle

\section{Introduction}\label{sec:intro}
Crowd-sourced developer forums like Stack Overflow (SO) are popular among developers.
The Stack Exchange network of sites that host Stack Overflow had 9+ billion page views from 100+ million users in 2019 alone.
Stack Overflow now hosts more than 50 million posts and is visited by 11 million users per day.
Questions related to coding challenges
often receive code examples as solutions. Stack Overflow contains
code snippets in 75\% of their answers ~\cite{SOcodesnippets}. The quality of
these examples and their direct reuse without modifications is a concern.
Previous studies have found that 9.8\% of 7,444 Stack Overflow accepted answers
contained at least one instance of a poor coding practice
~\cite{PythonVulnerabilities}, and
 Android code examples shared in
SO are reused in millions of popular Android app~\cite{cFISCHER}.

Collaborative editing is introduced in forums like SO to allow users to suggest ways to improve the shared content. Previous studies offer valuable insight on the security of SO C/C++ code examples and the effect of post revisions ~\cite{EmpiricalC++Study, SOcodesnippets, C/C++SO}.
However,
what happens when a vulnerable code example is revised?

In this paper, we attempt to answer this question by analyzing vulnerable crowd-shared C/C++ IoT code examples.
C/C++ is widely used in mission-critical systems and resource-constrained IoT devices.
The Internet of Things (IoT) is an internet connected system
of physical objects
("things")~\cite{InternetofThings}. Rapid developments in IoT have made it so
that an estimated 27.1 billion IoT devices will be connected by 2025
~\cite{StateofIoT}.
Increased demand for IoT in various use cases
consequently increases the importance of understanding
the unique challenges of IoT security ~\cite{SmartHomesSecurity}.

The prevalence of IoT devices in our everyday life and the ease of access to such computing resources make development using IoT devices widespread.
As such, recent research reports a growing number of IoT related posts in forums like SO~\cite{Uddin-IoTTopic-EMSE2021,uddin2021security}.
In addition, certain Stack Exchange sites are more specialized to certain fields
or technologies. For example, the Arduino and Raspberry Pi sites are designed
for discussion regarding development using these two tools, which are popular in
the field of IoT.

We study all IoT C/C++
code snippets shared in the SO, Arduino, and Raspberry Pi Stack Exchange sites.
We apply a static code vulnerability analysis tool (cppcheck) to check each
code example for vulnerabilities. For each identified vulnerable code example, we collect its
revision history from the three sites. We then check whether the vulnerability was introduced
pre or post-revision of a code example, and
whether certain revision types (e.g., code improvement) introduced or fixed vulnerabilities.

We find several severe vulnerabilities in the shared code snippets like \dv{788}{CWE 788 - Access of Memory Location After End of Buffer}. We also find that most snippets are not revised, and the majority of vulnerabilities are introduced pre code revisions. When revisions are made, we observe that they are more likely intended to improve the functionality of the code than to make simple changes. However, these revisions often have little effect on reducing the number of vulnerabilities in a code snippet. We conclude that the current editing system in the forums may be insufficient to help mitigate vulnerabilities in the shared code snippets.


\nd\textbf{Replication Package.}

\nd\url{https://github.com/disa-lab/esem2022-crowdeditcodevulnerable}

\section{Motivating Examples}\label{sec:motivation}
Our study was motivated by our observations of vulnerable C/C++ IoT code examples in the Stack Exchange forums.
Below, we show three examples of vulnerable C/C++ IoT code snippets that were modified during revisions.

\begin{figure}[t]
\includegraphics[width=\columnwidth]{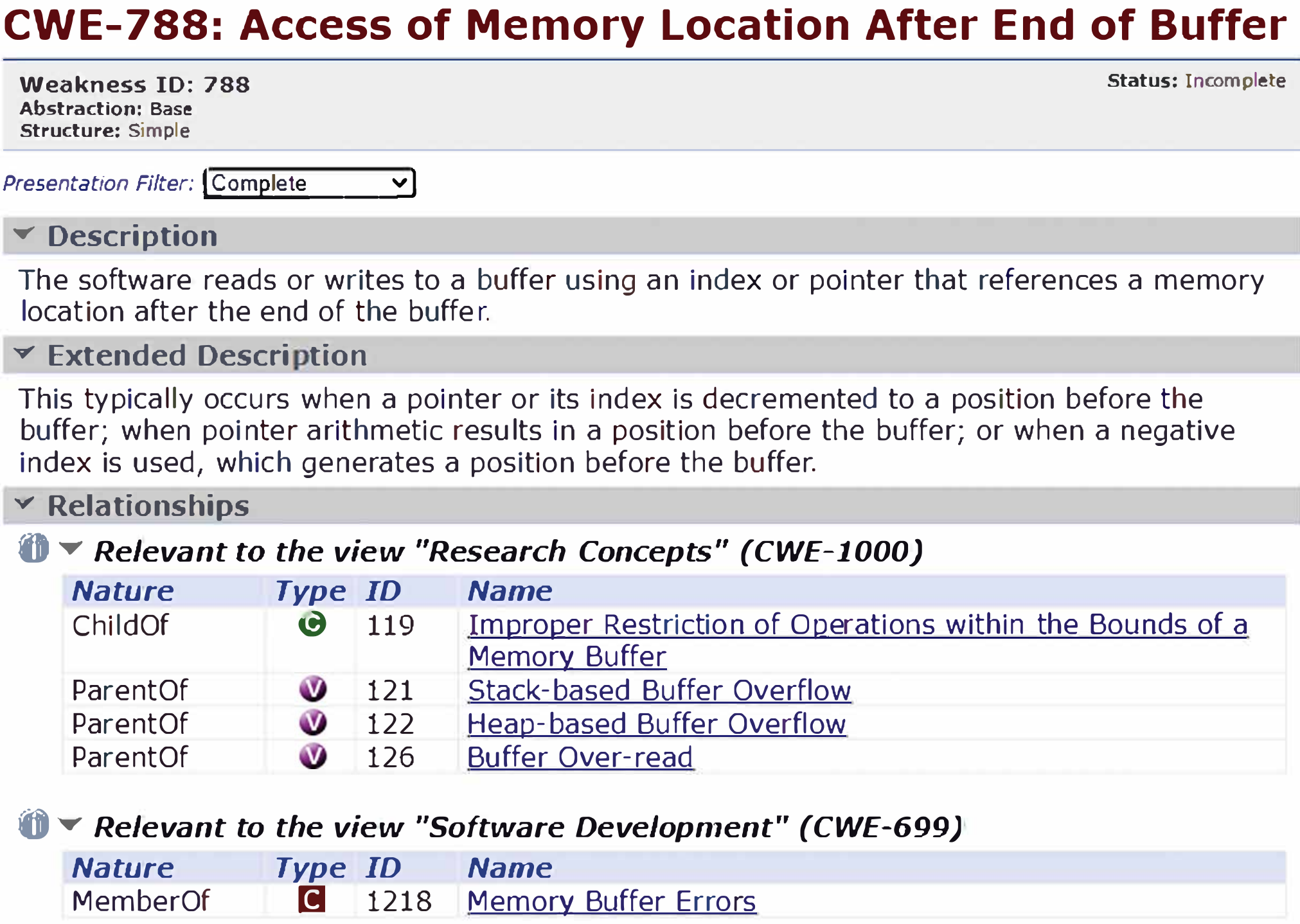}
\caption{Screenshot of \dv{788}{CWE 788} in the CWE Database}
\label{fig:cwe788-sc}
\end{figure}

\begin{figure}[t]
\centering
\includegraphics[width=\columnwidth]{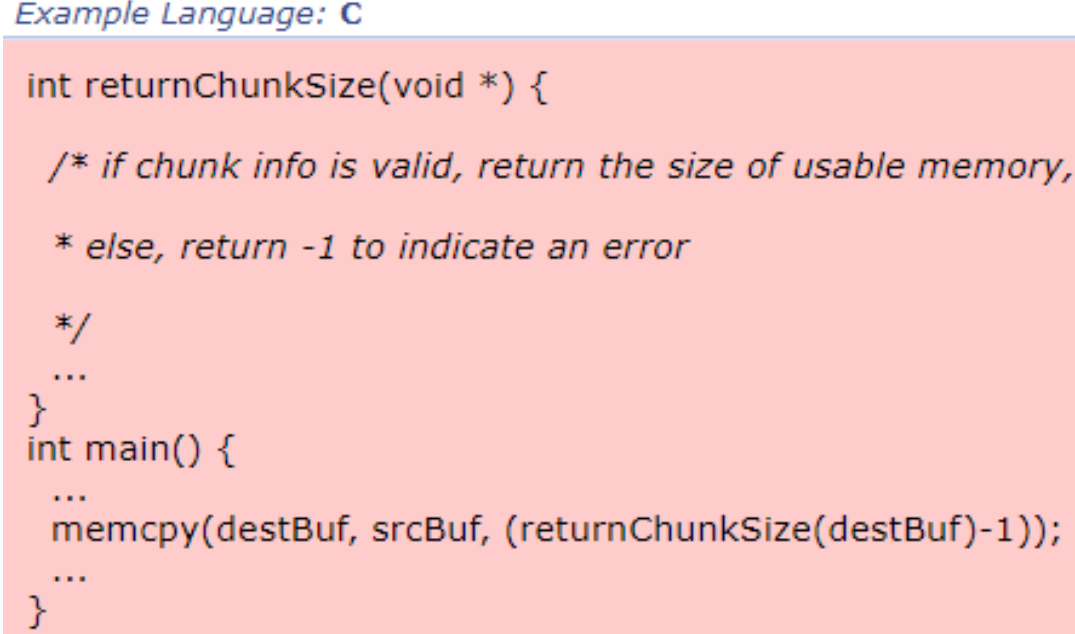}
\caption{An example vulnerable code pattern in online CWE database that shows a \dv{788}{CWE 788} vulnerability}
\label{fig:cwe788-ex}
\end{figure}

Listing 1 is an example of a SO code snippet that had vulnerabilities introduced in the
original version (\dv{788}{CWE 788 - Access of Memory Location After End of Buffer} at
lines 14 and 23), and then gained more vulnerabilities after it was revised, which introduced the same
vulnerability (\dv{788}{CWE 788}) in line 32. Common Weakness Enumeration (CWE) is a community based list of software
weaknesses maintained by the Mitre Corporation ir order to help catalog software vulnerabilities~\cite{CWEDef}. As of July 2021, there are a total of 924 weaknesses in CWE
Version 4.6 with 92 related to C/C++ ~\cite{CWEVersion,C_weakness,C++_weaknesses}. In \fig\ref{fig:cwe788-sc}, we show a screenshot of the official entry of \dv{788}{CWE 788} in the CWE
online database. The information contains the title of the weakness type, which is access of memory location after end of buffer.
The information also contains a description with common consequences and it's relationship to other CWE types.
Each entry also contains examples of the vulnerability with explanations as to why it is harmful. In Figure \ref{fig:cwe788-ex},
we see an example code snippet of \dv{788}{CWE 788} as provided in the online CWE database. This vulnerability can be exploited when
the C method \texttt{memcpy} is provided to copy a source memory location with a buffer.
In Listing 1, the revision in line 32 to the code example calls the \texttt{memcpy} method without checking the buffer size.
Therefore, instead of fixing the previous similar vulnerability in line 14, the revision has in fact made it worse.
\begin{figure}[t]
\includegraphics{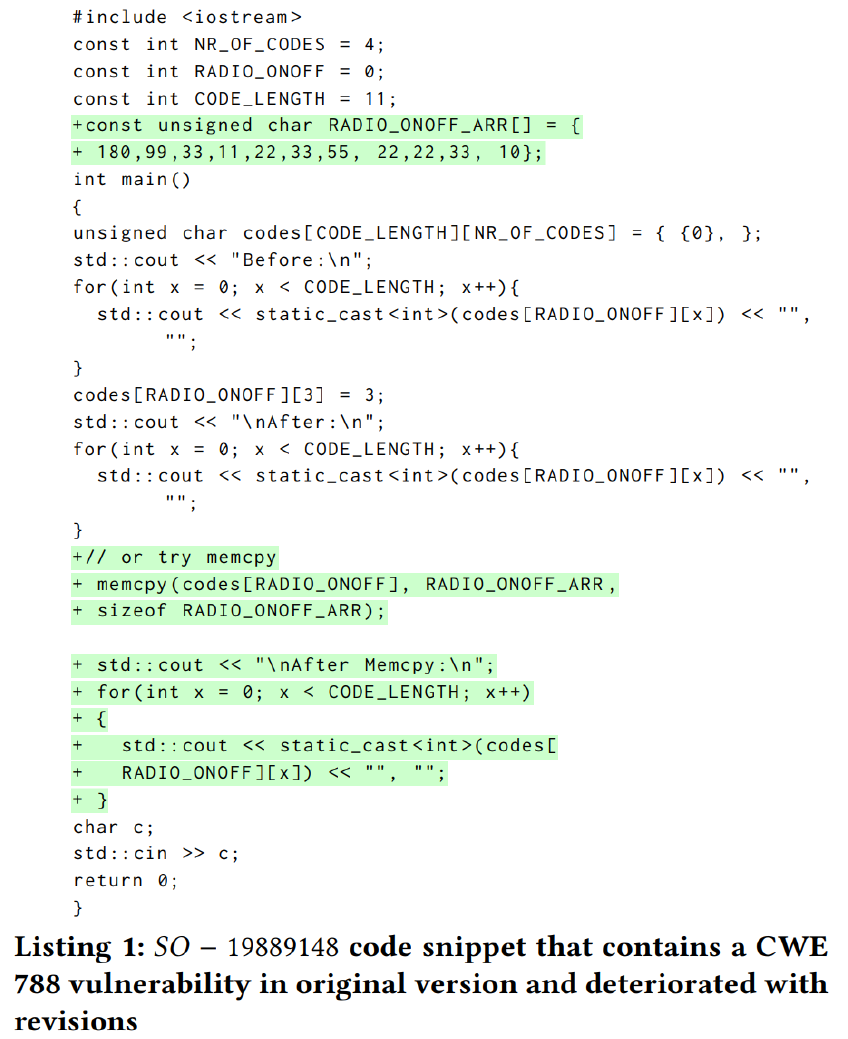}
\end{figure}
\begin{figure}[t]
\includegraphics{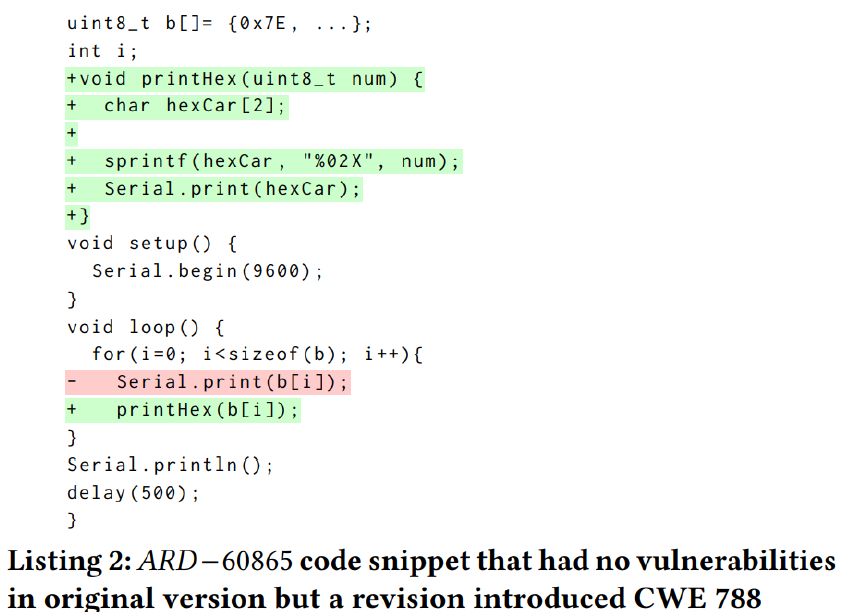}
\end{figure}
\begin{figure}[t]
\includegraphics{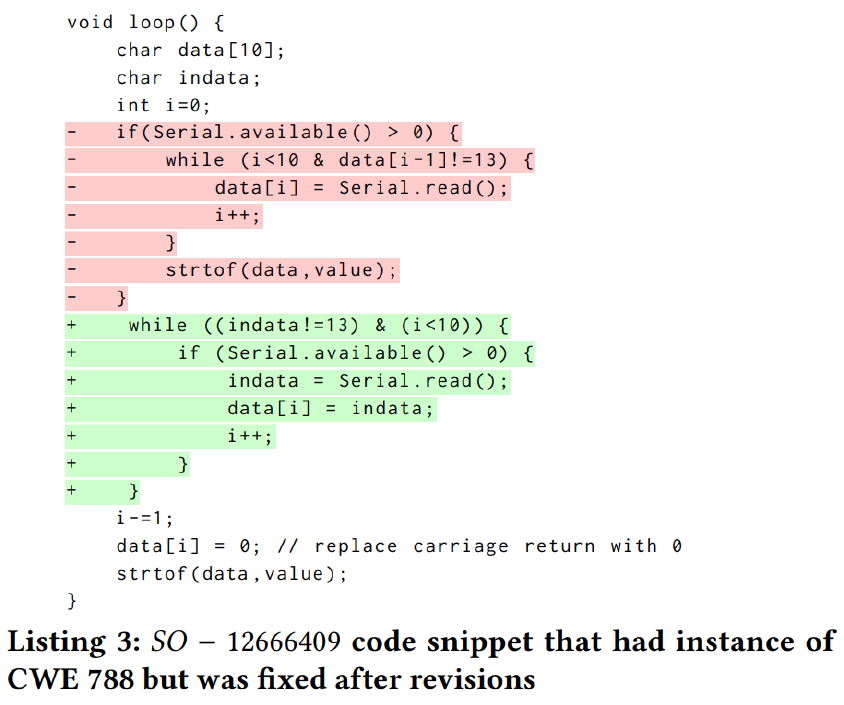}
\end{figure}

Some code snippets did not have any weaknesses in the original version posted by
the user, but then gained some as the user revised. The ARD code snippet in Listing 2 shows that
an instance of \dv{788}{CWE 788} was introduced in
line 10 because the user revised to explain how to print the
hexadecimal values of the array, but misused the sprintf function.
Finally, some code snippets were also improved by revisions. For example, in
a SO code snippet shown in Listing 3, the user removed an instance of \dv{788}{CWE 788 - Access of
Memory Location Before Start of Buffer} in line 6 by changing the way data is stored in
the data array. Therefore, revisions to code examples in online developer forums
can lead to improvements, leave the code unchanged, or even make it worse by introducing further weaknesses.
Our study aims to understand the proportion and types of vulnerable code examples that are mitigated through the revisions.
\section{Study Setup}\label{sec:study-setup}
We collect all IoT code snippets shared in three Stack Exchange sites
and preprocess those to identify and record their revisions.
\subsection{Data Collection}\label{sec:data-collection}
We study IoT code examples shared in the following three Stack Exchange sites: SO,
Arduino, Raspberry Pi. SO is the most popular Q\&A site for software developers of all kinds. The other two sites are
specifically setup to foster IoT-based discussions. We download the January 2022 data dump of each site, and then obtain all code examples present in answers
from Arduino and Raspberry Pi. For SO, we obtain code
examples from answers that belong to questions labeled as the 75 IoT-related tags from Uddin et
al.~\cite{Uddin-IoTTopic-EMSE2021}.

For SO, the question tags were checked to see if they contained the
keywords C or C++, and if they contained IoT related keywords. We observed however that
Arduino and Raspberry Pi questions did not have programming languages present in
their tags in most cases. In order to determine the language of the code
examples on these sites, we instead used the language
detection tool Guesslang which has an accuracy of 90\% according to it's
documentation \cite{guesslangdoc}. Then, we used a similar approach used in
previous studies to reject code snippets that only contained pseudo code by
ignoring snippets that contained less than the median SO line count of 5 lines
~\cite{SOTorrent,C/C++SO,SpottingCodeExamples}. We also
collected the entire version history of each of these code snippets. In total,
we obtain 13,170 code snippets from 10,248 posts with 19,018 post versions. A
breakdown of the collected snippets as well as their versions is shown in Table
\ref{fig:studied_sites_data_collection}.
 \begin{table}[t]
        \centering
        \caption{Statistics of each studied site}
        \begin{tabular}{lrrr}\toprule
        \textbf{Site Name} & \textbf{\#Code Snippets} & \textbf{\#Post Versions}\\
        \midrule
        Stack Overflow (SO) & 5,086 & 7,073\\
        Arduino & 6,906 & 9,393\\
        Raspberry Pi & 1,178 & 2,549 \\
        \textbf{Total} & \textbf{13,170} & \textbf{19,015}\\
        \bottomrule
        \end{tabular}
\label{fig:studied_sites_data_collection}
\end{table}

\subsection{Data Preprocessing}\label{sec:data-preprocessing}
The collected snippets were analyzed for weakness on the CWE
(Common Weakness Enumeration) List. To do this we used cppcheck version 2.4.1 released in March 2021, which is a
static code analyzer that supports various types of code checks in C and C++ code. It also allows for specific weaknesses to be suppressed. According to Zhang et al, cppcheck is
able to identify 59 out of the 90 code weaknesses that are related to C and C++
~\cite{C/C++SO}. Previous studies have found that cppcheck had just a 0.78 false
positive rate against a test suite of 650 common C/C++ bugs
~\cite{ArusoaieC/C++Analysis}. Zhang et al. found that 85 out of 100 CWE
instances detected by cppcheck were labelled as accurate with a strong agreement
among the study's authors  (Cohen's Kappa of 0.68) ~\cite{C/C++SO}.

While we analyzed the initial results of running the obtained code snippets
through cppcheck, we observed many instances of syntax errors. These errors are
likely to be automatically detected by code editors and removed by the
programmer, so we proceeded to ignore such errors in our analysis, similarly to
Zhang et al., who ignored 129,395 instances of
syntax errors in their initial observation of 154,198 CWE instances.
Other reported errors we noticed to be unfair to deem as a weakness are CWE types such as
\dv{563}{CWE 563 - Assignment to Variable without Use}.
Such errors are not important as users of Q/A sites often intend to answer specific questions in code examples with direct answers, not to provide complete solutions. We therefore suppress errors in cppcheck of this nature, which are summarized in Table \ref{fig:supressed_errors}.
\begin{table}[t]
        \centering
        \caption{Errors supressed in cppcheck}
         \resizebox{\columnwidth}{!}{%
        \begin{tabular}{p{3.3cm}|p{4.8cm}}\toprule

        \textbf{Criteria Name} & \textbf{Criteria Description}\\
        \midrule
        Syntax Error & Errors in the syntax of the code\\
        Unread Variable & Variable is assigned a value but never used\\
        Unused variable or unused stuct member & Variable or struct member is not assigned a value and then never used \\
        Unused private function & Private function is not called\\
        \bottomrule
        \end{tabular}%
        }
\label{fig:supressed_errors}
    \end{table}

\section{Study Results}\label{sec:study-results}
Our empirical study answers for research questions (RQ) to offer insights into the relationship between the vulnerability of IoT code snippets and their revisions in three Stack Exchange sites:
\begin{enumerate}[leftmargin=25pt]
  \item Were the vulnerabilities introduced through post revisions? (\sec\ref{sec:rq1})
  \item What are the different types of vulnerabilities found during the revisions? (\sec\ref{sec:rq2})
  \item Does the type of vulnerability differ depending on revision types? (\sec\ref{sec:rq3})
  \item Were the vulnerabilities introduced pre-edit mitigated via post revisions? (\sec\ref{sec:rq4})
\end{enumerate}

\subsection{RQ$_1$ Were the vulnerabilities introduced through post revisions?}\label{sec:rq1}
\subsubsection{Motivation}
Answers on Stack Exchange websites can be modified in order to improve it's
quality or to add further information. The goal of collaborative editing in the online forums is to foster
content quality, which can also include improving the quality of the shared code examples.
However, through these modifications, it is possible that
new security
vulnerabilities are introduced. An understanding of when these vulnerabilities are introduced could help us determine tools and guidelines by focusing on the timeline. For example, if most vulnerabilities are left unchanged post-edit, the collaborative editing systems needs to change.

\subsubsection{Approach} To determine the stage in which a a user may introduce a vulnerability in a code example, we first collect the entire version history of each post.
This version history was then manually analyzed for revisions to the code examples themselves. Therefore, edits made to the text portion of the answer were not considered as revisions. We utilized this method as minor text changes such as typo fixes will have no affect on the vulnerabilities present in the code segment.
After obtaining all code snippet revision history, we analyzed snippets that contained multiple versions for vulnerabilities in order to determine which version the error was introduced. Using Cppcheck to detect CWE types, we compared the results from the revised snippets with the initial ones.
\subsubsection{Results}
\begin{figure}[h]
    \centering
\includegraphics{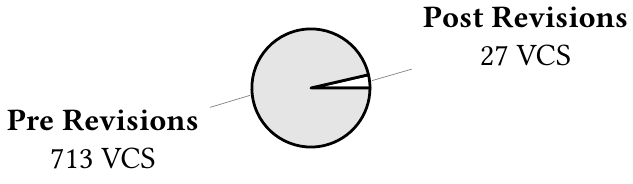}
	\caption{Distribution of vulnerable code snippets (VCS) by the stage when their vulnerabilities were introduced (pre vs post revision of a code snippet)}
	\label{fig:pre-post-pie}
\end{figure}

Out of our studied 13,170 code examples across the three Stack Exchange sites, we found 740 code snippets flagged as vulnerable by the cppcheck tool.
We then tracked the revisions of each code snippet, and identified when the vulnerability was introduced.
We find that that the vast majority of vulnerabilities are introduced before
edits are made to code snippets, i.e., when the code snippet was first shared. As shown in Figure \ref{fig:pre-post-pie}, 713
out of the 740 vulnerable code snippets were either not changed, or contained
CWE instances before revisions were made.

Table \tbl\ref{tab:pre-post-table} shows the distribution of the vulnerable code snippets by the stage when the vulnerability
was first found in the corresponding code snippet (pre or post-revisions). We see that only a few of the vulnerable code snippets in each site
were introduced post-edit, i.e., most of the snippets contained a vulnerability when the code examples were first shared.
Out of the three sites, 95.8\% of the vulnerable code snippets in SO had the vulnerabilities introduced pre-edit. Arduino had 96.6\% of the vulnerable code snippets introduced during pre-edit and
3.4\% introduced during post-edit. Finally, Raspberry Pi had
no vulnerabilities introduced after revisions.
\begin{table}[h]
  \centering
  \caption{Distribution of VCS introduced pre and post edit across the studied sites }
\begin{tabular}{@{}c@{}}
\includegraphics{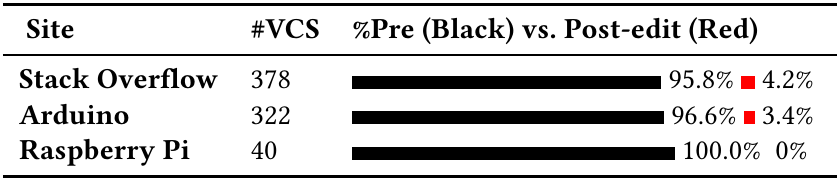}
\end{tabular}
    \label{tab:pre-post-table}
\end{table}%

\begin{figure}[h]
\centering
\includegraphics[scale=0.43]{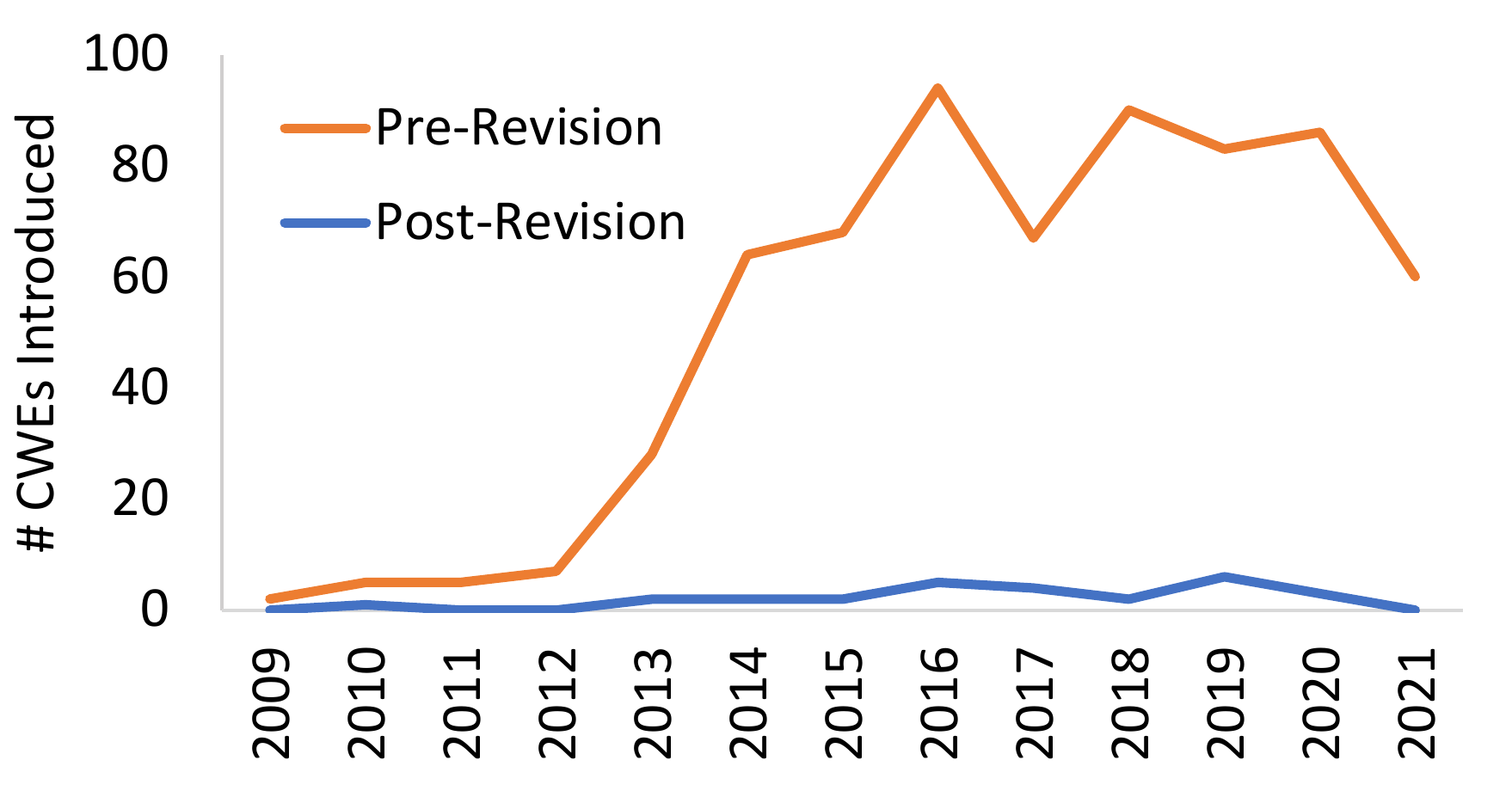}
\caption{Evolution of \#CWE instances introduced pre revisions vs. post revisions}
\label{fig:trends}
\end{figure}

Listing 4 shows a vulnerability that was not
present in original version but was introduced once the user made a revision in
an attempt to improve their code. In this case, an instance of \dv{398}{CWE 398 - 7PK Code
Quality} in line 13 was introduced after the inclusion of a for loop that uses a
variable (loop) with the same name as another function.
An example of a vulnerability that was introduced before revisions and was never fixed during revisions can be found in SO post (\soa{27918518}), where an instance of \dv{401}{CWE 401 - Missing
Release of Memory after Effective Lifetime} is present in line 5. \fig\ref{fig:trends} shows
the evolution of the number of CWE introduced pre and post revisions from 2009 to 2021. We find that although the number of CWEs introduced pre revisions dramatically increased between 2012 and 2016, it has been on a slight decline since. Furthermore, the number of new CWEs introduced after a revision has been made has remained relatively consistent and small.

\begin{figure}[h]
\centering
\includegraphics[scale=0.5]{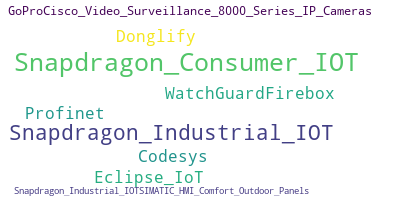}
\caption{Common IoT Devices affected by the CWEs}
\label{fig:wordcloud}
\end{figure}

We also find the IoT devices affected by the CWE types detected in the code snippets by analyzing the CVE descriptions of the CWE types. As shown in Fig \ref{fig:wordcloud}, Snapdragon Consumer IoT and Snapdragon Industrial IoT are the most common affected IoT devices, followed by other devices such as GoPro cameras.


\begin{figure}[t]
\includegraphics{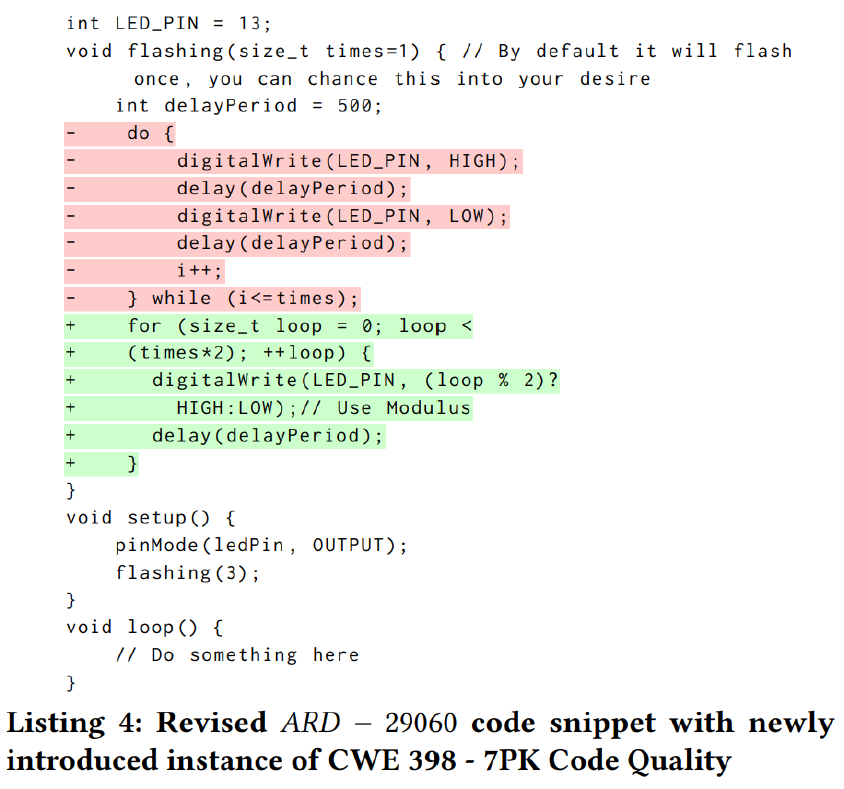}
\end{figure}

\vskip\baselineskip
\noindent\includegraphics[width=\columnwidth]{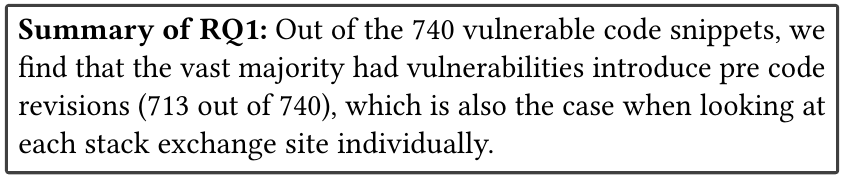}
\vskip\baselineskip


\subsection{RQ$_2$ What are the different types of vulnerabilities found during the revisions?}\label{sec:rq2}
\subsubsection{Motivation}
We analyze the vulnerabilities and CWE types present in the Stack Exchange code snippets to better understand their overall security. We also aim to determine if revisions add or remove certain types of vulnerabilities in order to gain more insight on the effect of code revisions.
\subsubsection{Approach}
We use the output generated by cppcheck to determine the types
of CWEs present, and the CWE database for their descriptions and characteristics. Using these
characteristics, we try to understand the root cause behind the weaknesses in
order to determine a categorization. This is done by both of the paper's
authors. We then check how revisions impacted the distribution of vulnerable code snippets under each category.
\subsubsection{Results}
\begin{figure}[t]
    \centering
	\includegraphics{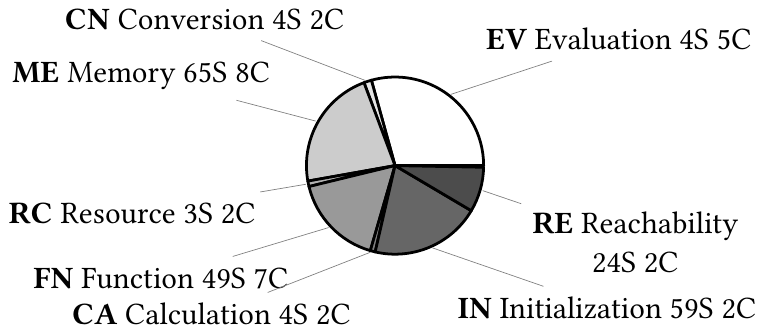}
	\caption{Distribution of the observed 30 CWE Types by the eight weakness categories (S = weak code snippet, C = CWE)}
	\label{fig:vul-cat-dist}
\end{figure}

Across the three stack exchange sites, we observe a total of 31 CWE types present in
the 740 vulnerable code snippets (total of 1221 CWE instances), with many code snippets containing more than one CWE type.
The most frequently occurring CWE type we find is \dv{398}{CWE 398 - 7PK Code
Quality}, which
was present in 520 code snippets. This CWE type however is not a distinct
vulnerability, but instead indicates that the code snippet is of poor quality.
We categorize the remaining 30 distinct CWE types based on their characteristics
into the following 8 weakness categories: Evaluation -  incorrect logic and comparisons, Memory - insufficient memory management, Function - improperly called and designed functions ,
Initialization - improperly initialized variables, Reachability - unreachable or undefined code, Resource - mismanagement of a program's resources, Conversion - incorrect conversions of variables to different types, and Calculation - improper or incorrect calculations.
In Figure
\ref{fig:vul-cat-dist} we observe that code snippets are more likely to contain
CWE types that are evaluation, memory, and initialization related.
\begin{table}[t]
  \centering
  \caption{Distribution of CWE Types by number of VCS (= Vulnerable Code Snippet). Pre = Black bar, Post = Red bar}
\begin{tabular}{@{}c@{}}
\includegraphics[width=\columnwidth]{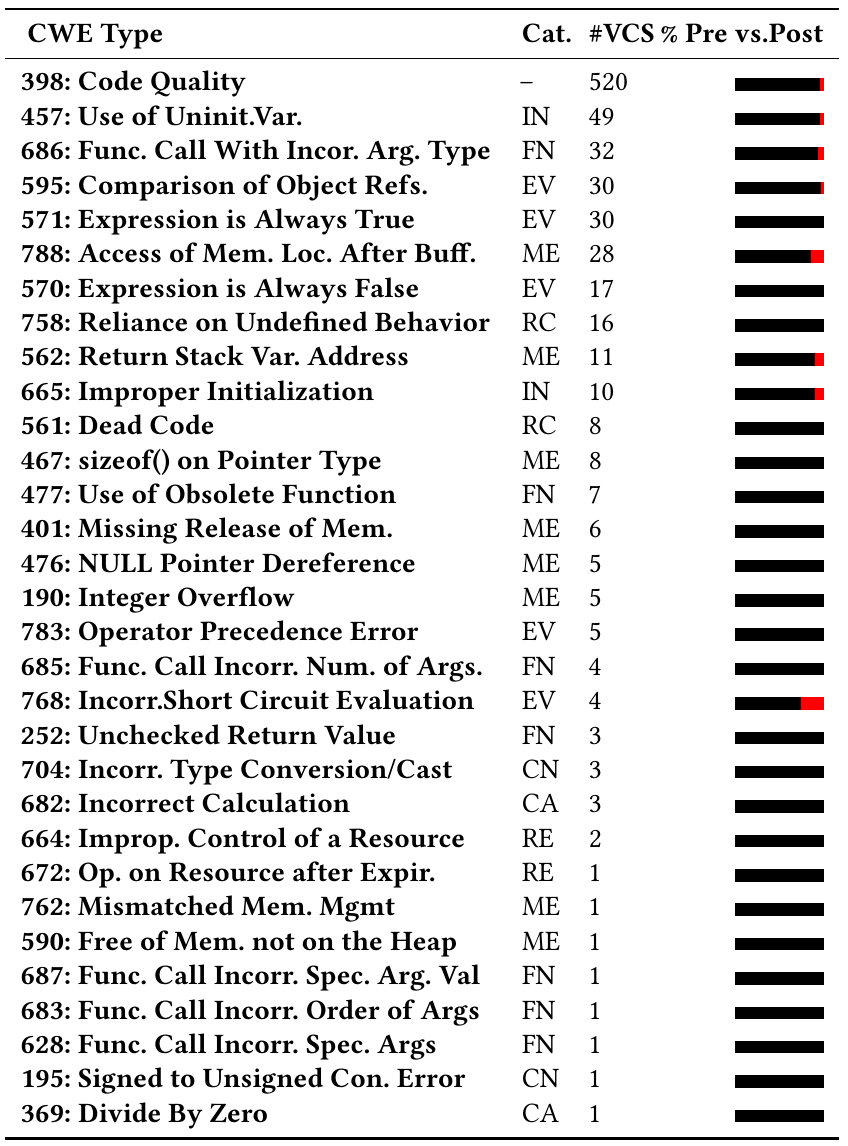}
\end{tabular}
    \label{tab:cwe-dist-table}
\end{table}%
\begin{table}[t]
  \centering
  \caption{Distribution of code snippets with weaknesses introduced pre and post edits, and code snippets improved after revisions by weakness category}
\begin{tabular}{@{}c@{}}
\includegraphics[width=\columnwidth]{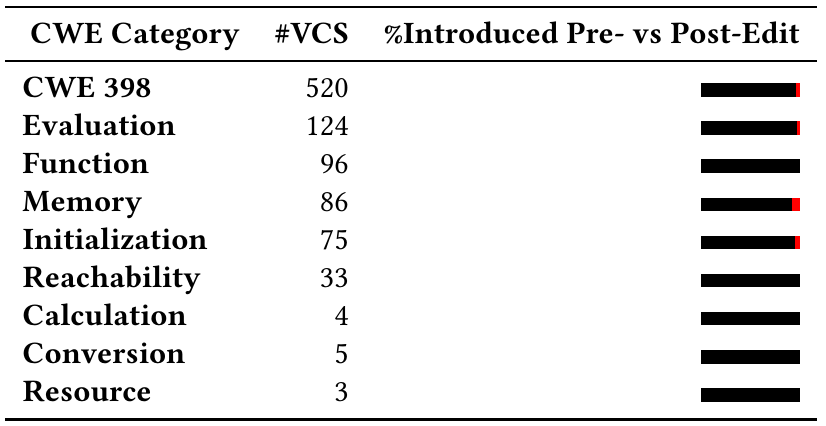}
\end{tabular}
    \label{tab:cat-table}
\end{table}%

Among the 30 distinct CWE types shown in Table \ref{tab:cwe-dist-table}, we find
that \dv{457}{CWE 457 - Use of Uninitialized Variable} is the most frequently occurring
(detected in 49 out of 294 code snippets), followed by \dv{686}{CWE 686 - Function Call
With Incorrect Argument Type} (32 out of 294).

Table \ref{tab:cat-table} displays the proportion of vulnerable code snippets in each weakness category that have weaknesses introduced pre or post-edit. We find that across all categories and types, vulnerabilities are introduced more frequently before revisions are made than after, and for most categories there are no vulnerable code snippets that contain weaknesses introduced post revisions. We also observe that memory related CWE instances are more likely to be introduced by code revisions compared to evaluation related weaknesses or instances of \dv{398}{CWE 398}.

When looking at code snippets that contained vulnerabilities before revisions, but then saw a reduction in CWE instances after, we find that only 3 weakness categories (evaluation, memory, and initialization) and \dv{398}{CWE 398} contained such snippets.

\vskip\baselineskip
\noindent\includegraphics[width=\columnwidth]{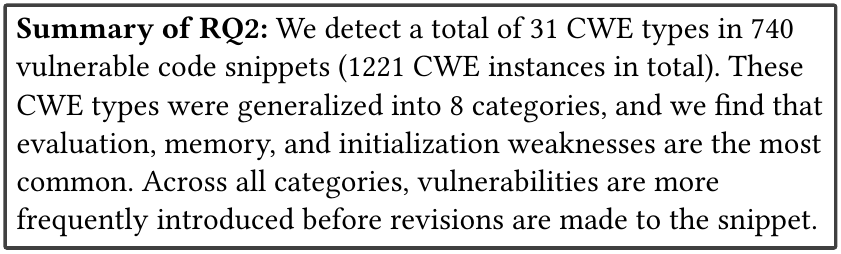}
\vskip\baselineskip


\subsection{RQ$_3$ Does the type of vulnerability differ depending on revision types?}\label{sec:rq3}
\subsubsection{Motivation}
Users revise their answers for a variety of reasons. We aim to better understand the behaviour of these users by analyzing the types of revisions that are more commonly made when editing code snippets.
\subsubsection{Approach}
We manually analyze code snippets that have at least one code revision by looking at the nature of the edit. Then following work by Wang et al.~\cite{SOcodesnippets}, we label each revision with four revision types: code correction, code formatting, code improvement, and code removal/addition.
\subsubsection{Results}
\begin{figure}[h]
    \centering
	\includegraphics{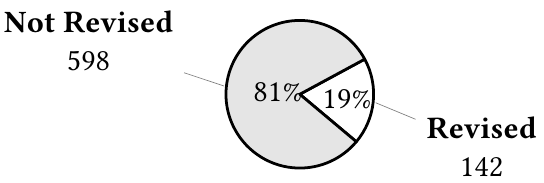}
	\caption{Number of code snippets revised and not revised}
	\label{fig:rev-vs-nonrev}
\end{figure}

\begin{figure}[h]
    \centering
	\includegraphics{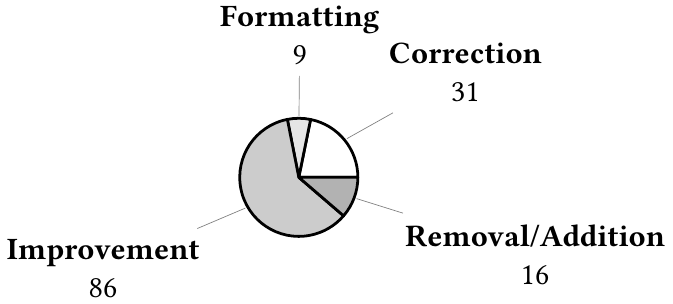}
	\caption{Distribution of revised vulnerable code snippets (VCS) by their revision type}
	\label{fig:rev-type-dist}
\end{figure}

An overview of the 4 revision types we label the revised code snippets as is shown in Table \ref{tab:revision_def}. Overall we find that 142 out of the 740 vulnerable code snippets are revised, as shown in Figure \ref{fig:rev-vs-nonrev}. Out of the 598 vulnerable code snippets that did not have any code revisions, 327 originate from posts that were never revised, meaning the user did not make any changes to the original version of the post. The remaining 271 code snippets were not revised, but rather were part of posts that were only textually edited. For example, users edit posts to correct typos, clarify their work, or to include references and links to other resources.
For the 142 code snippets that were revised, we observe that users more frequently make revisions with the intention of improving the functionality of the example to better answer the question (86 out of 142 snippets, 60.6\%), and to make corrections in the code such as fixing syntax errors and typos (31 out of 142, 21.8\%). As shown in Figure \ref{fig:rev-type-dist}, the least common type of edit we observe in the code snippets with revisions is formatting changes, such has adding white space through indentation (9 out of 142 code snippets).

\begin{table}[t]
  \centering
  \caption{Revision types and their definitions}
  \label{tab:revision_def}
    \resizebox{\columnwidth}{!}{%
    \begin{tabular}{p{2.2cm}p{6cm}}
    \toprule{}
    \textbf{Revision Type} & \textbf{Definition}\\
    \midrule
    \textbf{Code Correction CR} & Changes to the syntax, and typo changes in comments and variable/function names. \\
    \midrule
    {\textbf{Code Formatting FM}} & Adding white space or newlines, improving the formatting/readability of the code. \\
    \midrule
    {\textbf{Code Improvement IP}} & Functionality or performance changes to the code, includes changes to logical expressions, calculations, and the types of variables or return values. \\
    \midrule
    {\textbf{Code Removal / Addition RA}} & Removing or adding code segments, for example adding a new function or class.\\
    \bottomrule
    \end{tabular}%
  }
\end{table}%

\begin{table}[t]
  \centering
  \caption{Distribution of revised code snippets by their revision type and type of weakness (Revision Types from \tbl\ref{tab:revision_def} CR = Black, FM = Blue, IP = Magenta, RA = Green). \#TV = Total Vulnerable Code Snippet VCS, \#RV = Revised VCS. Cat. = CWE categories from \fig\ref{fig:vul-cat-dist}}
  \label{tab:revision_type_dist}
\begin{tabular}{@{}c@{}}
\includegraphics[width=\columnwidth]{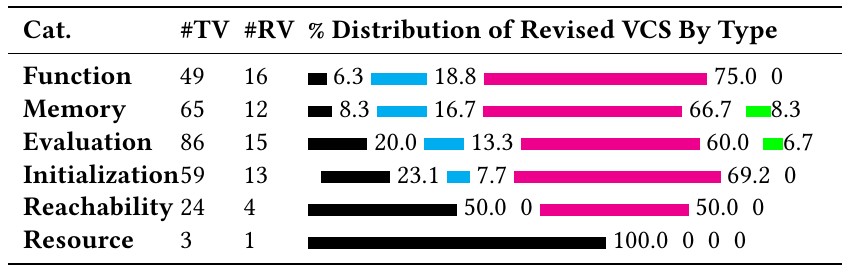}
\end{tabular}
\label{tab:dist-revtype-vcs}
\end{table}%

We then further analyze the relationship between the CWE type present in a revised code snippet and its revision type. Table \ref{tab:revision_type_dist} shows that across the 6/8 weakness categories that contain vulnerable code snippets that were revised, the most common revision type is code improvement. We also find that code correction revisions are more common in snippets with evaluation and initialization weaknesses (3 and 4 snippets respectively), and code removal/addition revisions are more common in snippets with memory weaknesses (5 snippets).
Overall, there is a statistically significant difference between the number of vulnerable code snippets and the number of revised vulnerable code snippets across the six categories (Mann Whitney U test $p = 0.02$).

\vskip\baselineskip
\noindent\includegraphics[width=\columnwidth]{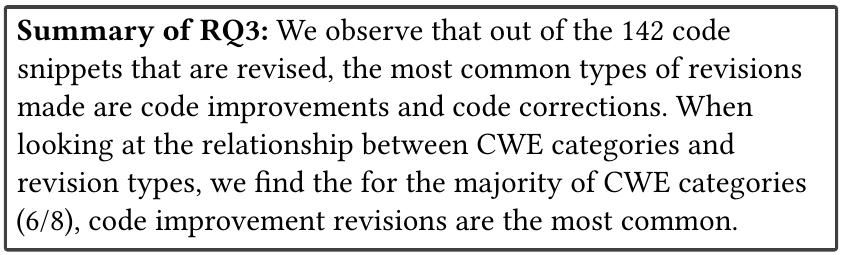}
\vskip\baselineskip

\subsection{RQ$_4$ Were the vulnerabilities introduced pre-edit mitigated via post revisions?}\label{sec:rq4}
\subsubsection{Motivation}
Revisions made by stack exchange users can improve code snippets by fixing
errors and reducing vulnerabilities, or they can further deteriorate an already
vulnerable snippet by adding new weaknesses. Revisions may also leave
vulnerabilities unchanged. It is thus important to better understand the impact of code revisions on reducing vulnerabilities.
\subsubsection{Approach}
Using the version history of the obtained code snippets, we first determine the number of code revisions for each code snippet.
For the snippets that have at least one revision, we run all versions through cppcheck to determine if the number of vulnerabilities stayed the same, decreased, or increased.
\subsubsection{Results}
\begin{figure}[h]
\centering
\includegraphics[scale=0.32]{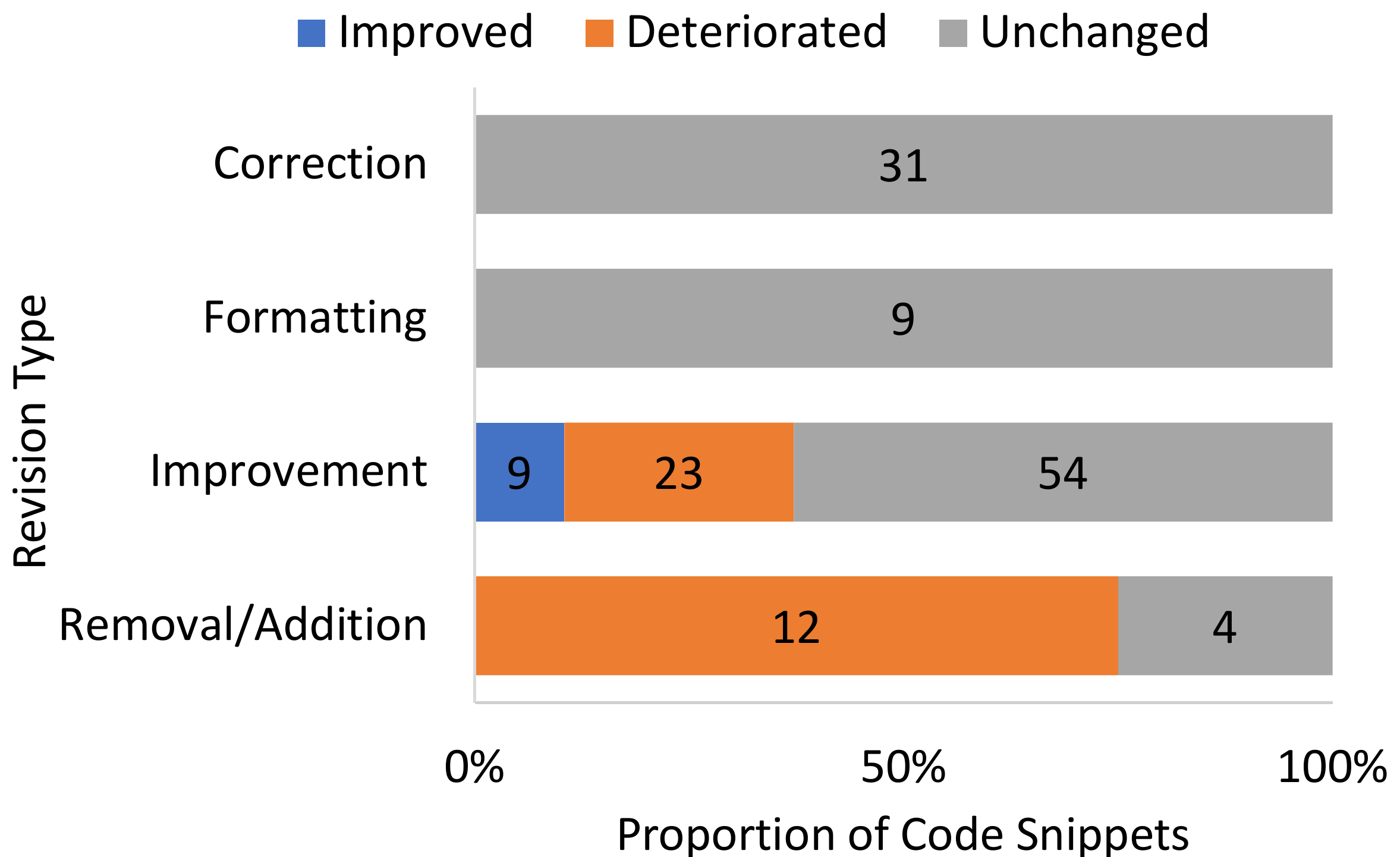}
\caption{Distribution of revision effect (improved, deteriorated, or unchanged) by revision type}
\label{fig:rev-type-effect}
\end{figure}

We look at whether certain revision types are correlated more closely with particular revision outcomes, such as reducing the number of vulnerabilities in the snippet or introducing new ones. As shown in Fig \ref{fig:rev-type-effect}, for the 40 revised code snippets that contained correction and formatting revisions, all did not experience any change in vulnerabilities. We further find that all 9 code snippets that were improved due to revisions were revised with the intent to improve the code, such as Listing 3. However, a large number of code snippets with improvement related revisions also experienced deterioration or no change in the number of vulnerabilities (23 and 54 code snippets respectively). Finally, we notice that code snippets with revisions related to the removal or addition of large segments of code were more likely to deteriorate (12 out of 16 snippets). This is likely due to the fact that the addition of large code segments increases the chance of introducing a weakness. For example, in the following Stack Overflow answer (\soa{52560467}), the user revised the snippet by adding the check\_relay function and the large switch statement in the enable\_relay function. Although these changes may have been made with the intention of better answering the question, they ultimately introduced an instance of \dv{398}{CWE 398} in line 35 due to the scope of the variable state.

\begin{figure}[h]
\centering
\includegraphics[scale=0.28]{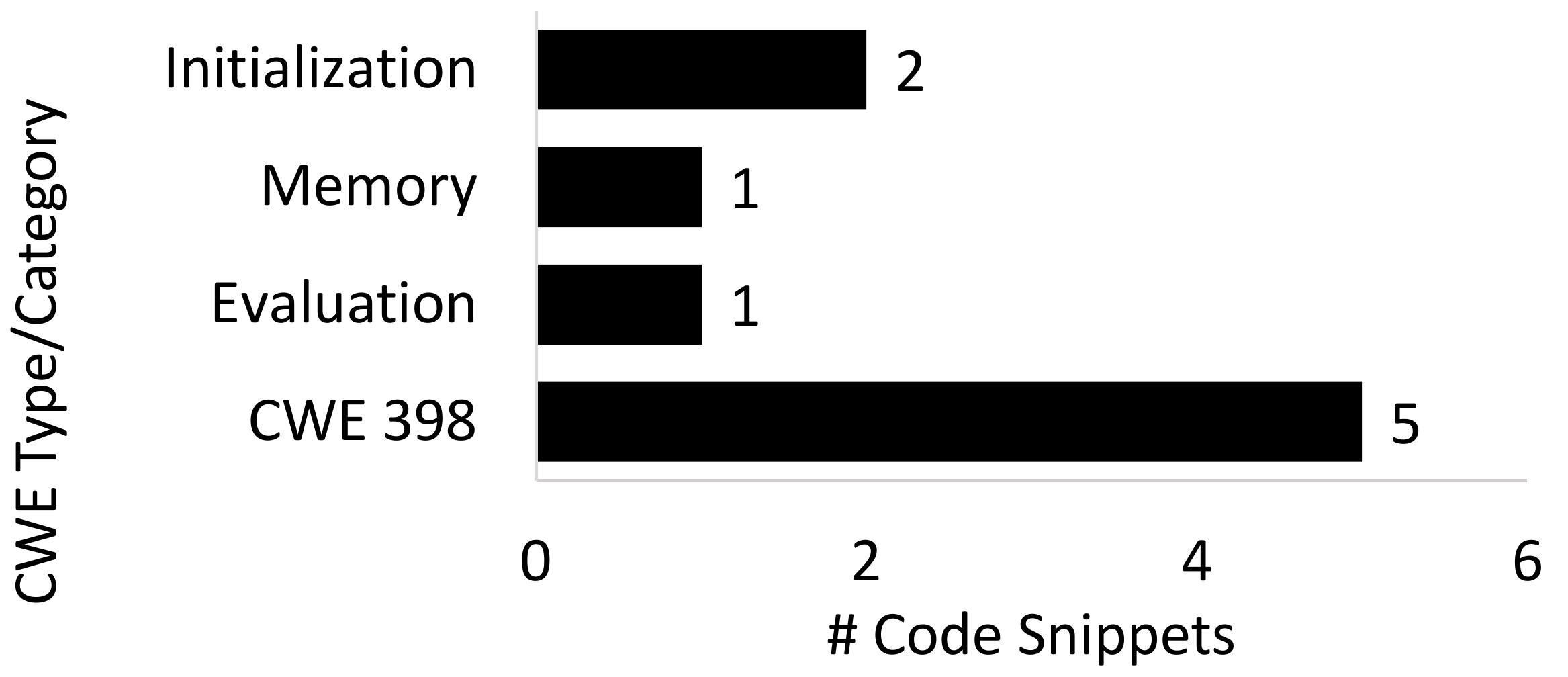}
\caption{Distribution of code snippets that were improved due to revisions by CWE type or CWE category}
\label{fig:improved}
\end{figure}

Overall, only a small number of code snippets experienced an improvement due to revisions (9 in total). As shown in Figure \ref{fig:improved}, those with initialization related weaknesses and instances of CWE 398 contained the most (2 out of 59, and 5 out of 520 respectively).

\begin{figure}[h]
\centering
\includegraphics[scale=0.28]{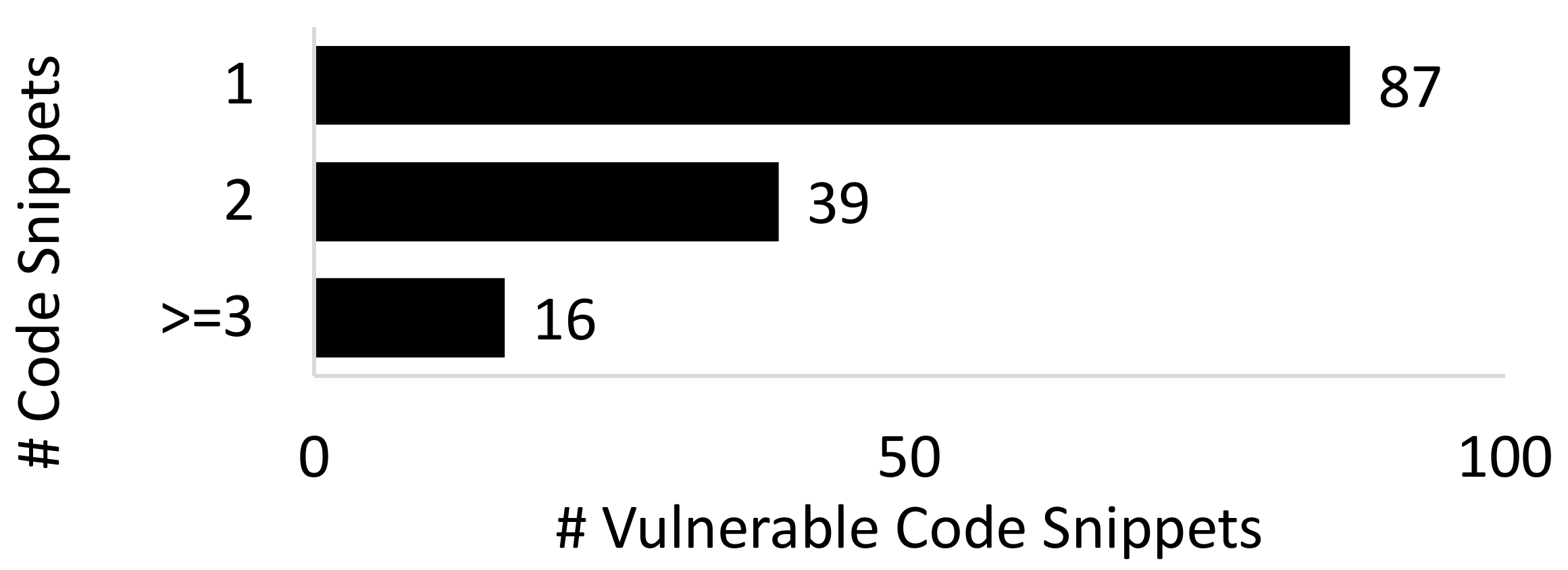}
\caption{Distribution of the number of code revisions}
\label{fig:dist-code-rev}
\end{figure}

\begin{figure}[h]
    \centering
	\includegraphics{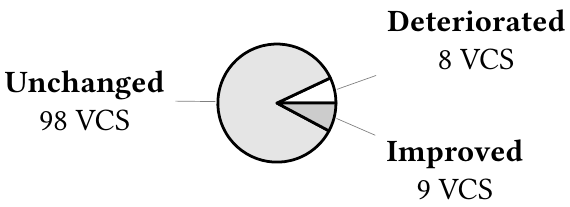}
	\caption{Distribution of revision effect in snippets with pre-existing vulnerabilities}
	\label{fig:revision-effect}
\end{figure}

In total, 115 code snippets with pre-existing vulnerabilities experienced one or more revisions. In Figure \ref{fig:dist-code-rev}, we observe that the majority of all revised code snippets (in total 142) were revised just once.
However, we also observe that the effect of revisions on these code snippets is minimal. Figure \ref{fig:revision-effect} shows the vast majority of these code snippets with pre-existing vulnerabilities that were revised did not experience a decrease or increase in vulnerabilities (98 out of 115). We also find that in some code snippets (9 out of 115), revisions removed vulnerabilities that existed in previous versions. However, in a similar number of code snippets (8 out of 115), vulnerabilities were introduced by revisions.

Stack Overflow answer \soa{25827521 } is an example of a snippet that was revised once but did not improve in terms of removing its weakness. By comparing the final version of the code to the original, we observe that the user likely intended to improve the functionality and correctness of the code. We see they changed the function parameters to void instead of leaving it empty to ensure no arguments could be passed, and they also attempted error handling by first checking if read\_AT\_string did not return NULL. However, they did not address the instance of \dv{562}{CWE 562 - Return of Stack Variable Address} detected by cppheck in line 36 where the return value was a pointer to a local variable, which is invalid in C++. This is an example of a revision that went beyond non-functional changes such as adding comments or fixing typos, but was still not effective at removing the weakness in the snippet.

\begin{figure}[t]
    \centering
	\includegraphics{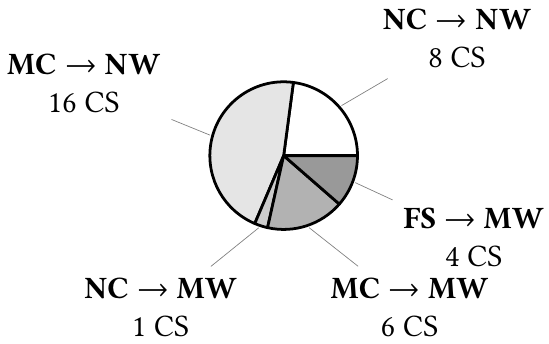}
	\caption{Distribution of deterioration type (M = Modified, N = New, C = Code, W = Weakness. FS = Fixed Syntax.)}
	\label{fig:det-type}
\end{figure}

We analyze the 31 deteriorated code snippets for similarities in the changes made to the code, and the subsequent CWE instances that were introduced. In total, we find 4 distinct deterioration types : (1) New component added + introduced new weakness, (2) Modified existing component + introduced new weakness, (3) New component added + introduced more existing weaknesses, and (4) Modified existing component + introduced more existing weaknesses. We further find that a few deteriorations occurred due to the user fixing a syntax error in the code, which allowed for more weaknesses to be detected by cppcheck (5). The distribution of these 5 types is shown in Figure \ref{fig:det-type}, where we see that the most common type of deterioration was caused by users modifying an existing component in the code (such as a function, variable, or class), and introducing new CWE instances that were not present before.

\vskip\baselineskip
\noindent\includegraphics[width=\columnwidth]{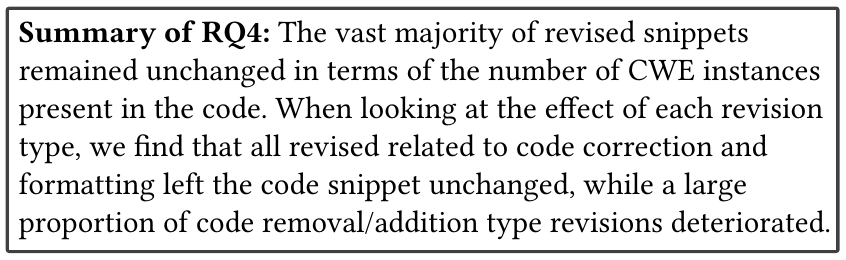}
\vskip\baselineskip

\section{Discussion}
Our study findings can guide the following stakeholders: IoT developers, IoT vendors, Forum designers, and IoT researchers and educators. We discuss the implications below.

\begin{figure}[h]
\centering
\includegraphics[scale=0.38]{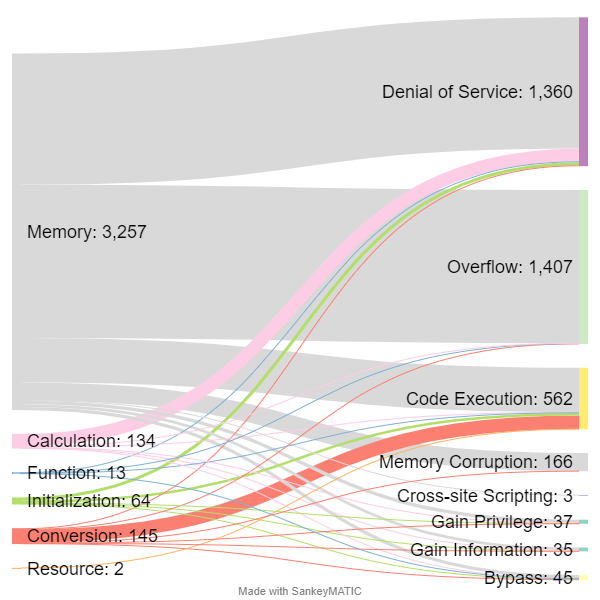}
\caption{Mapping of CWE categories to CVE Types. Width of each line represents the number of CVEs per CWEs.}
\label{fig:flow-diagram}
\end{figure}
\nd\textbf{{IoT Developers:}} In \fig\ref{fig:flow-diagram}, we show the number of CVEs reported in the NVD database per the six vulnerability types we observed
in our IoT code snippets (in pre + post revisions). The reported CVEs are grouped under eight types in the cvedetails.com site: 
(1) Denial of service,
(2) Overflow,
(3) Code execution,
(4) Memory corruption,
(5) Cross-site scripting,
(6) Gain privilege,
(7) Gain information, and
(8) Bypass.
 This \fig\ref{fig:flow-diagram} shows that the six vulnerability types in the IoT code snippets can cause all the eight types
of critical security issues in the IoT devices, if the vulnerable code snippets are reused from the Stack Exchange sites. Therefore,
IoT developers need to stay aware of the potential security problems in such shared code snippets in the developer forums. Our study findings like
the catalog of CWEs and the associated vulnerable code snippets can be used by the IoT developers to learn about such pitfalls. Such knowledge can
be useful for the developers in multiple phases. First, when they share any IoT code examples in the forums. Second, when they revise a shared code example.
Third, when they reuse a shared code example. In all phases, they can check the code for vulnerability by consulting our catalog of CWEs. Given that
revisions from others rarely fix such a vulnerable code snippets, it is important for the IoT developers to practice such quality assurance of the code snippets,
even after the code snippet is revised by other users in the forums.

\noindent\textbf{\underline{IoT Vendors:}} In \fig\ref{fig:flow-diagram}, we showed that the six vulnerability type we observed in the shared IoT code snippets can
be mapped to eight CVE types in NVD database. In the NVD database, each reported CVE is categorized into the severity types:  Critical (C), High (H),
Medium (M), and Low (L).  In \tbl\ref{fig:cvss_distribution_by_CWE}, we show the distribution of the CVEs by four severity types. Overall, we see that all
six vulnerability categories in our observed IoT code snippets can contribute to many critical and highly severe security vulnerabilities in the IoT devices.
In \fig\ref{fig:wordcloud}, we showed that such affected IoT devices can belong to many vendors (e.g., Cisco, Snapdragon, etc.). Therefore, IoT device
and SDK vendors can work together to make the devices more resilient. One solution would be to incorporate automated security testing
tool into the IoT SDKs and devices. The
IoT vendors cannot rely much on the collaborative editing system in the crowd-sourced developer forums to improve the code snippets.
\begin{table}[h]
  \centering
  \caption{Distribution  of the 12 CWEs with mapped CVEs in the cvedetails.com database. Each colored bar under CVSS score category denotes a severity category (Black = Low, Cyan = Medium, Magenta = High, Red = Critical)}
  \label{fig:cvss_distribution_by_CWE}
\begin{tabular}{@{}c@{}}
\includegraphics[width=\columnwidth]{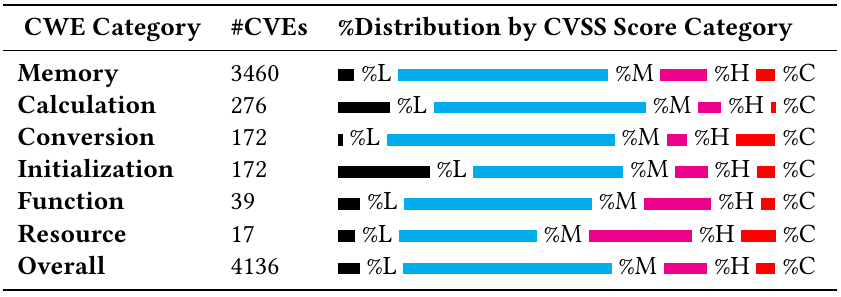}
\end{tabular}
\end{table}%

\begin{table}[h]
  \centering
  \caption{Distribution of weakness categories by revision type. Each colored bar denotes a different category.
  (Black = Function, Green = Memory, Magenta = Evaluation, Red = Initialization, Cyan = Reachability, Orange = Resource)}
  \label{fig:cve_type_distribution}
\begin{tabular}{@{}c@{}}
\includegraphics[width=\columnwidth]{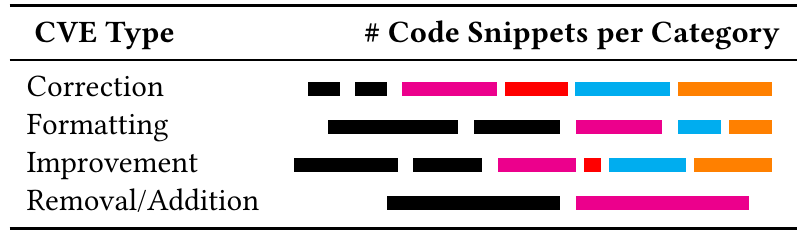}
\end{tabular}
%
\end{table}
\begin{figure}[h]
\centering
\includegraphics[scale=0.32]{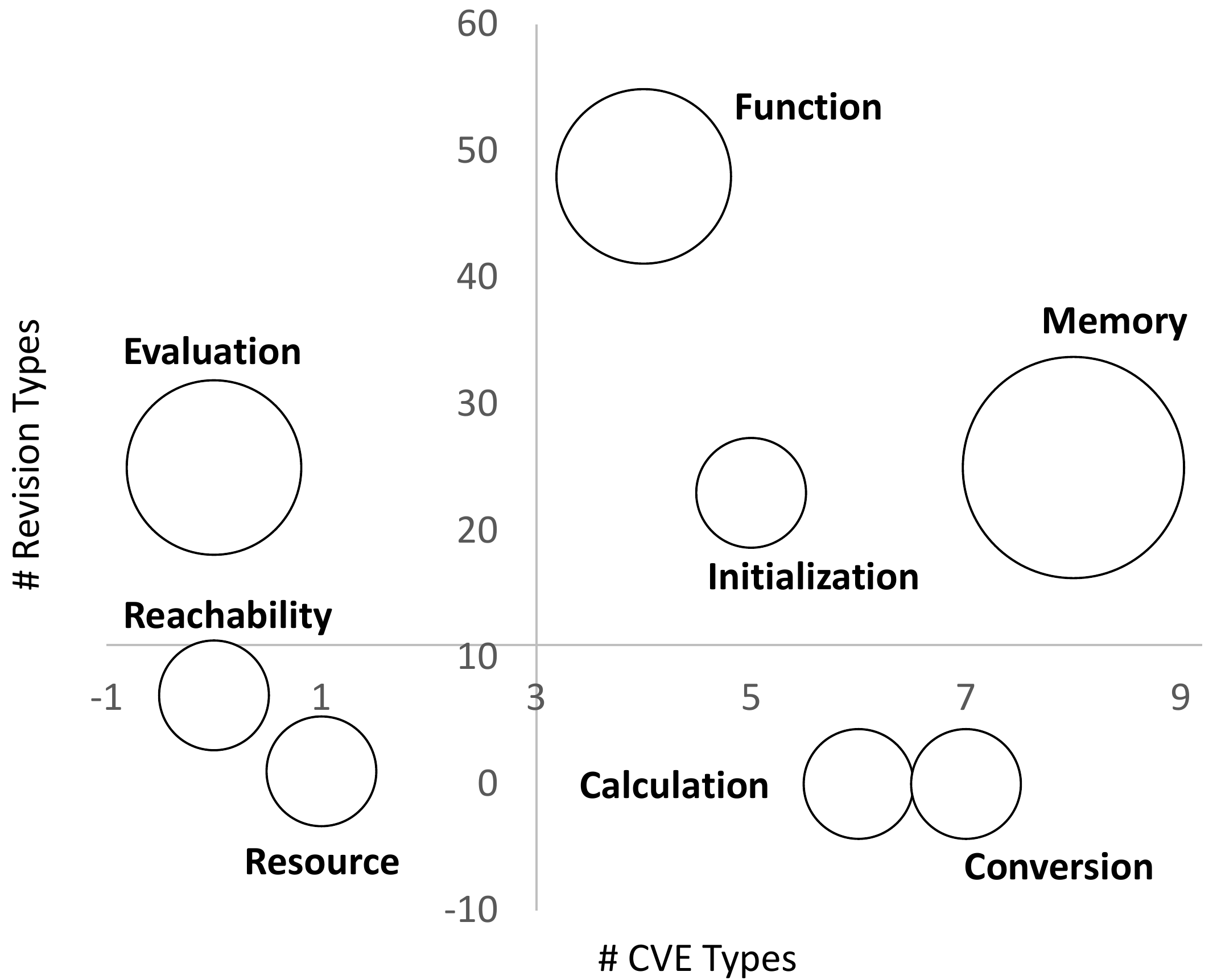}
\caption{The tradeoff between the revisions and the vulnerabilities introduced towards the IoT code snippets}
\label{fig:bubble}
\vspace{-2mm}
\end{figure}

\noindent\textbf{\underline{Forum Designers:}} In \tbl\ref{fig:cve_type_distribution}, we show the distribution of our observed six vulnerability types across the four types of code revisions
we observed, i.e, whether and how a revision type introduced a vulnerability type in the revised code. We find that correction to a code introduced all the six
types of vulnerability categories. Intuitively, this observation goes against the general assumption that correction to a code should have fixed more vulnerability.
The statistics from \tbl\ref{fig:cve_type_distribution} can inform forum designers of the fact that the current mechanisms (e.g., the four revision types) to support
collaborative editing in the online forums are insufficient to help mitigate security vulnerabilities in the shared code.
In \fig\ref{fig:bubble}, we reinforce this observation by showing a bubble chart as a tradeoff between the number of revisions made to our studied vulnerable code snippets
and the number of distinct CVEs types that were introduced during the revisions. The size of each bubble corresponds to the number of total CVEs reported against the CWE IDs found
per CWE category.  \fig\ref{fig:bubble} shows that the most number of CVEs reported for the `Memory' category weaknesses, and code snippets with such weaknesses in our
dataset were revised considerably with little or no success in fixing. On the other hand, code snippets belonging to `Conversion' type weaknesses were revised almost never.
Overall, there is a positive correlation between the number of revisions made to a code snippet and the number of vulnerabilities found in the code snippet.
The findings call for a redesign in the collaborative editing process in the forums by the designers of the sites, e.g., to facilitate the incorporation
of security guidelines into the editing process by providing automated recommendations.

\noindent\textbf{\underline{IoT Researchers and Educators:}} Our findings offer a grim picture on the effectiveness of collaborative editing to help
mitigate security vulnerabilities in developer forums. IoT researchers can join hands with both the forum designers
and the IoT vendors to conduct research on the better design of the collaborative editing and to incorporate security validation framework into the IoT devices.
Our findings reinforce the recent worries on software supply chain attacks~\cite{NIST-SoftwareSupplyChainAttack}
by showing that security vulnerabilities are prevalent in online developer forums and they are mostly left unaddressed during revisions.
One way to help IoT developers during the sharing of code snippets in online forums is to educate them with on-demand documentation about the security
issues by consulting security patterns and the vulnerabilities reported in the NVD database.
The IoT security educators can join hands with the forum designers to produce such documentation, which can also offer new directions to the
current approaches that utilize online developer forums to create and/or improve software documentation~\cite{Uddin-HowAPIDocumentationFails-IEEESW2015,Uddin-OpinerAPIUsageScenario-IST2020,Uddin-OpinerReviewAlgo-ASE2017,Uddin-OpinionValue-TSE2019,Uddin-ResolveAPIMentionSO-TechReport2017,Chakraborty-NewLangSupportSO-IST2021}.
This is important given studies that IoT developers do indeed consult about security issues in online developer forums~\cite{uddin2021security}, but they also
face difficulty to get answers to their questions~\cite{Uddin-IoTTopic-EMSE2021}.








\section{Threats to Validity}
\textbf{Internal validity} threats relate to author bias in deciding which
weaknesses to ignore. We noticed
that some claimed weaknesses in cppcheck were not accurate. We
addressed this issue by suppressing certain CWE types in
Cppcheck like CWE 563 (Assignment to Variable without Use) and by following
a previous study~\cite{C/C++SO} to suppress syntax errors
and to ignore code snippets with less than 5 lines.
In addition, the categorization of the 28
distinct CWE types into 8 weakness categories was done by
both authors.
\textbf{Construct validity} threats relate to errors that may have occurred during
the data collection.
To determine if a SO
post was related to IoT, we used tags from existing studies~\cite{uddin2021security,Uddin-IoTTopic-EMSE2021}.
For Arduino and Raspberry, we made the assumption that all posts would be
related to IoT. Another threat is our use of
the language detection tool guesslang to identify C/C++ code snippets.
Guesslang has been used in previous studies to specifically
detect C/C++ code and has a validity rate of 90\%~\cite{guesslangdoc}.
\textbf{External validity} threats relate to how our findings can be generalized to
the nature of IoT posts on online Q/A sites as a whole. We focused
on vulnerabilities in Stack Exchange answers. This is because
code snippets found in answers are meant to be `solutions' and are more likely
to be copied and used by developers. We also focus
our study on strictly C and C++ code snippets due to their popularity in IoT
development.

\section{Related Work}
To the best of our understanding, the C/C++ code examples shared in Stack Overflow were subject to two empirical
studies recently, first by Verdi et al.~\cite{EmpiricalC++Study} and then by Zhang et al.~\cite{C/C++SO}.
Our study differs from the two studies as follows.
\begin{enumerate}[leftmargin=10pt]
  \item While both Verdi et al.~\cite{EmpiricalC++Study} and Zhang et al.~\cite{C/C++SO}
  analyze C/C++ code examples in general, we focused on IoT C/C++ code examples. While both
  analyzed only SO code examples,
  we studied data from three sites: SO, Arduino, and Raspberry Pi.
  \item While Verdi et al.~\cite{EmpiricalC++Study} analyzed C/C++ code for weakness,
  we studied revisions to the vulnerable C/C++ IoT code.
  \item Unlike Verdi et al.~\cite{C/C++SO} and Zhang et al.~\cite{C/C++SO},
  we studied the vulnerability types and their relationships with different revision types.
  Our fous is to learn whether and how revisions could help mitigate the observed vulnerability types.
\end{enumerate}

We observed some similarities and differences between our study results and above two papers. First, in all SO C/C++ code snippets,
Zhang et al., found 32 CWE types~\cite{C/C++SO} while
Verdi et al. found 31 CWE types~\cite{EmpiricalC++Study}. We found 28 distinct CWE types identified in the IoT C/C++ code snippets. Zhang et
al., who similar to us used cppcheck to automatically detect CWE instances, found
that 1.82\% of their collected code snippets (11,748 out of 646,716) contained
weaknesses. However, we observed a slightly higher proportion for IoT code examples as having at least one CWE
(6.4\%). Verdi et al. manually reviewed all of their
72,483 code snippets. They
found vulnerabilities in 99 (i.e., 0.14\%) of their SO code snippets.

Other related work can be broadly divided into \textbf{Studies} and \textbf{Techniques} to understand and mitigate IoT security issues.

\textbf{{Studies}} investigated underlying middleware solutions~\cite{Chaqfeh-ChallengesMiddlewareIoT-2012}, big data
analytics~\cite{Marjani-IoTDataAnalytics-IEEEAccess2017}, and the
design of secure protocols and
techniques~\cite{Fuqaha-IoTSurveyTechnologiesApplications-IEEECST2015,Khan-IoTSecurityReview-FGCS2018,Zhang-IoTSecurityChallenge-SOCA2014}
and their applications on diverse domains (e.g.,
eHealth~\cite{Minoli-IoTSecurityForEHealth-CHASE2017}).
SO posts have been previously studied for insecure python vulnerabilities ~\cite{PythonVulnerabilities},
topics discussed by IoT developers ~\cite{Uddin-IoTTopic-EMSE2021,uddin2021security,Mandal-IoTSecurityAspectSO-IST2022}, big
data~\cite{Bagherzadeh2019} and chatbot issues~\cite{abdellatifchallenges}.
\textbf{{Techniques}} and safety measures are
studied in Soteria \cite{Celik-IoTSafetySecurityAnalysis-USENIX2018},
IoTGuard~\cite{Celik-IoTDynamicEnforcementOfSecurity-NDSS2019}.  IoT devices
can be easy targets for cyber threats~\cite{Zhang-IoTSecurityChallenge-SOCA2014,Frustaci-IoTSecurityEvaluation-IEEEIoTJournal2017}. Encryption and
secure hashing technologies~\cite{Tedeschi-LikeSecureIoTCommunications-IEEEIoT2020}. Many authorization
techniques for IoT are proposed like SmartAuth~\cite{YuanTian-APIBot-ASE2017}.
For smart home security, IoT security techniques are proposed like Piano~\cite{Gong-IoTPIANO-ICDCS2017},
smart authentication~\cite{He-RethinkIoTAccessControl-USENIX2018}, and cross-App
Interference threat mitigation~\cite{Chi-SmartHomeCrossAppInference-DSN2020}.
Attacks on Zigbee, an IEEE specification used to
support interoperability can make IoT devices
vulnerable~\cite{Ronen-IoTNuclearZigbeeChainReaction-SP2017}.
We are not aware of any studies that checked
the effectiveness of revisions to help mitigate vulnerabilities in online shared code.

\section{Conclusion}
We analyzed code examples from the Stack Overflow, Arduino, and Raspberry Stack Exchange sites. We focused on analyzing weaknesses by analyzing their revision history. We found a total of 31 CWE types present in 740 code snippets. We observed that the vast majority of vulnerabilities are introduced pre code revisions (713 out of 740). When snippets are revised, the number of vulnerabilities that are present in that snippets are likely to not change. Our results indicate the collaborative editing in the forums do not help mitigate the code vulnerabilities. Our future work will focus developing techniques to incorporate security recommendations into the collaborative editing process.

\balance
\bibliographystyle{ACM-Reference-Format}
\bibliography{sample-base}

\end{document}